\documentstyle[11pt,newpasp,twoside,psfig]{article}

\index{summary}
\index{instructions}
\index{template}

\def\edcomment#1{\iffalse\marginpar{\raggedright\sl#1\/}\else\relax\fi}
\reversemarginpar

\begin{document}
\title{Pulsar observations at Mt. Pleasant}
\author{D. R. Lewis$^{1,2}$, R. G. Dodson$^1$, P. D. Ramsdale$^1$ and P. M. McCulloch$^1$}
\affil{$^1$University of Tasmania, Department of Mathematics and Physics, GPO Box 252-21, Hobart Tasmania 7001 AUSTRALIA}
\affil{$^2$Australia Telescope National Facility, CSIRO, PO Box 76, Epping NSW 1710 AUSTRALIA}

\begin{abstract}  Two daily pulsar monitoring programs are progressing at the Mount Pleasant Observatory, Hobart, Tasmania, Australia.  A new system involving the 26-metre radio telescope monitors 10 young pulsars daily and is focussed on near-real-time glitch finding.  This will allow Target of Opportunity observations to measure post-glitch heating of the neutron star surface (Helfand, Gotthelf, \& Halpern 2001).

The 14-metre continues its 21st year of daily monitoring of the Vela pulsar with a recent comprehensive frontend upgrade.  This is prior to an upgrade of the backend equipment currently in progress.  The 14-metre observed the most recent glitch of the Vela pulsar in January 2000 to the highest time resolution of any glitch and revealed a particularly short-term decay component (Dodson, McCulloch, \& Lewis 2002).  This decay component will provide constraints to the nature of the coupling of the stellar crust to the liquid interior.
\end{abstract}

\section{Introduction}  Mt. Pleasant Observatory is devoting resources to observe glitches at high resolution.  Two telescopes with specifically designed equipment are automated for continuous monitoring and glitch reporting of multiple pulsars.

The 14-metre radio telescope monitors the Vela pulsar for over 18~hours each day, and when set, PSR~B1641-45.  Both pulsars are known to be young, active and to glitch.  The Vela pulsar glitches approximately every 1000~days.  The 14-metre monitors this pulsar for the specific purpose of timing during and around glitch events.

Three uncooled receivers are mounted at the prime focus of the 14-metre to allow continuous dispersion measure determination.  Two are stacked disk, dual polarisation with central frequencies of 635~MHz and 990~MHz additional to a right hand circular helix at 1391~MHz.  Bandwidths are 250~kHz, 800~kHz and 2~MHz respectively, limiting pulse broadening from interstellar dispersion to less than 1\%.  The output is folded for two minutes giving an integrated pulse profile of 1344~pulses.  The backend to the 990~MHz receiver also has incoherent dedispersion across 8 adjacent channels allowing a study of individual pulses. 

A new addition is the daily monitoring of 10 other young, active pulsars with the 26-metre radio telescope.  A cryogenic dual polarisation 20~cm receiver with system temperature of 60~K centred at 1522~MHz is fed to a backend recording total intensity from a filterbank of 24, 2.5~MHz bandwidth channels.

\section{Vela's glitches}  The single pulse data from the latest Vela glitch showed a previously unobservable fast decay term (Dodson et al. 2002).  The post-spin-up short-term decay is below the resolution of previously reported two minute folds.  Using such data would derive a glitch epoch that is too early.
These high time resolution studies show a fast decay that can be modelled by a core response (Figure 1).  It will provide constraints on the nature of the coupling of the stellar crust to the liquid interior (Lewis, Link, \& Dodson in preparation).  The current data shows a need to improve the signal-to-noise ratio of the system in order to examine this phenomenon in greater detail.

The 14-metre time of arrival data archive has been analysed with improved telescope position, pulsar position and recent revised proper motion (Legge 2002).  A full list of glitch separation and size has been produced with a wealth of information on minor glitch events to be compared with similar work (Cordes, Downs, \& Krause-Polstorff 1988).

The post-glitch behaviour shows there is no simple pattern in the major observed glitches 5 to 13 in Vela (Figure 2). The preliminary result confirms a low braking index (Lyne et al. 1996), but is extremely sensitive to where after the glitches the second derivative of frequency is fitted.  Vela is important to discussions of true to characteristic age of pulsars (Gaensler \& Frail 2000; Gaensler et al. 2000; Kaspi et al. 2001).
   
\section{Futures}  The unique devotion of the 14-metre at Mt.~Pleasant will be taken advantage of in the near future by completing a sampling and recording system to coherently dedisperse a 32~MHz (both polarisations) instantaneous bandpass on the 635~MHz centred system.  The coherent dedispersion system is an adaption of the Mets\"ahovi disk-based VLBI recorder (Ritakari \& Mujunen 2002) with sufficient disk space for two hours of archived data. The data will be stored on the looping disk for post glitch analysis along side the 990~MHz system that at present automatically alerts us to a glitch event.  The new backend will increase the detected bandwidth at 635~MHz by a factor of 150.  The decrease in the frontend system temperature, will improve the signal-to-noise by at least an order of magnitude over the current 990~MHz single pulse system.  The receivers, amplifiers, filters, power supplies, backend and software of the 14-metre have been refined and reinstalled, improving the capabilities of the telescope.

Improvements of increasing bandwidth, observing time, reliability, number of pulsars monitored, a second larger telescope and reduced system noise are all aimed at observing more glitches, more effectively.  The motivation is to improve timing resolution to gain more information of the short time-scale spin-up events similar to that observed in 2000.

An important consequence of the upgrades to the Mt.~Pleasant automated system is the turn-around time for glitch reporting is now hours (day and night).  During the 2000 event, this allowed a Target-of-Opportunity observation on the Chandra X-ray observatory (Helfand et al. 2001).  Future events will be in conjunction with more formal ToO requests to find heating of the stellar crust.

\section {Summary}  Glitches must be observed with higher timing resolution to study the short-term decay in more detail.  This provides an insight to the interior physics of neutron stars and Equations of State.
As glitches are not precisely predictable, a continuous monitoring program is required.  The new and improved programs in Hobart increase the chances of observing glitches in real-time.

Continuous monitoring when combined with glitch alerting allows observation of the post-glitch heating of the crust.  Post-glitch heating is believed to be resultant from the energy deposition during a glitch (vanRiper, Epstein, \& Guy 1991; Hirano et al. 1997).

X-ray observations of the neutron star as close as possible to the glitch epoch are required to show all heat dissipation time-scales.  A study of the amount of heat energy dissipated over time will allow further insights into neutron star structure (Hirano et al. 1997; vanRiper et al. 1991).

\pagebreak
\begin{figure}
  \psfig{file=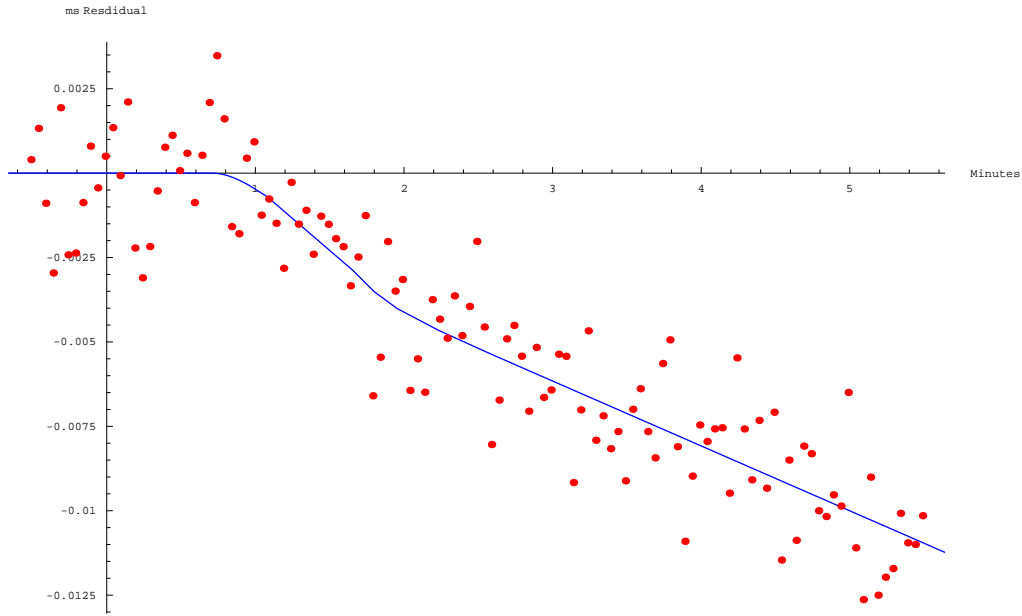,width=13.5cm}
  \caption{Example core response modelling with 3-second folds from the January 2000 Vela glitch} 
  \label{fig:main image}
\end{figure}

\begin{figure}
  \psfig{file=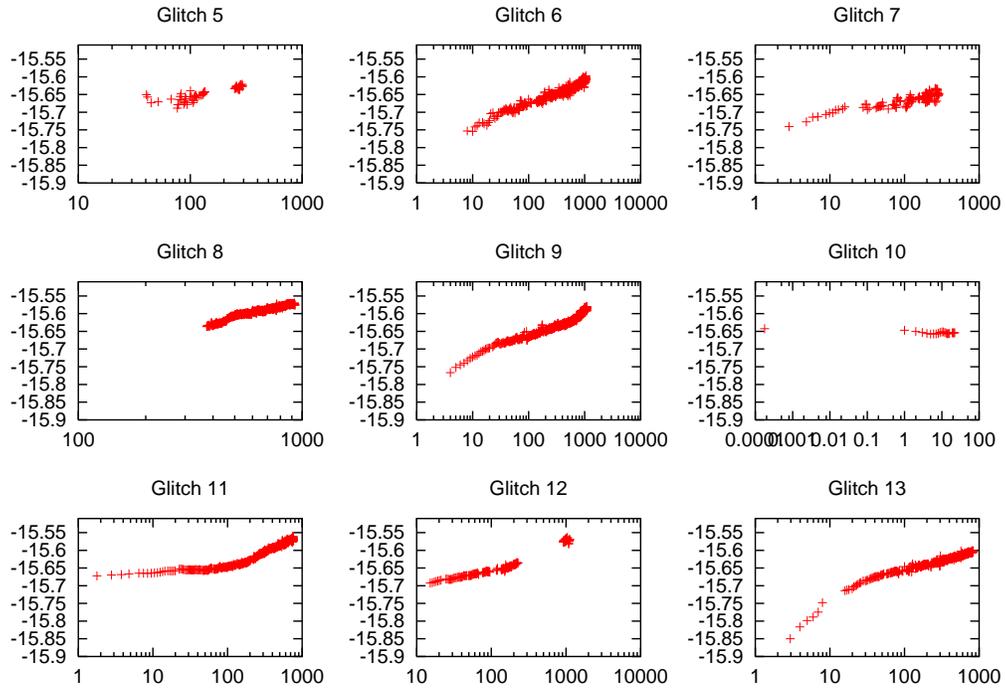,width=13.5cm,angle=270}
  \caption{20 years of Vela glitches - Frequency derivative vs. Log days} 
  \label{fig:main image}
\end{figure}

\end{document}